**Short Paper**

# PRESENT: An Android-Based Class Attendance Monitoring System Using Face Recognition Technology

Djoanna Marie Vasquez Salac
Batangas State University ARASOF-Nasugbu
dsalac17@gmail.com

**Abstract**

*Purpose* – The study aimed to develop an Android-Based Class Attendance Monitoring Application using Face Recognition to make attendance checking and monitoring easier and faster. In addition, the system has also integrated the use of SMS technology to notify parents/guardians whenever the students attended a particular subject or class. In addition, the study also intended to evaluate the developed application in terms of functionality, usability, reliability, and portability.

*Method* – The researcher utilized the incremental model. After the development phase, the application was evaluated by seventeen (17) faculty members from the College of Engineering and Computing Sciences. A validated evaluation questionnaire was used to rate the level of acceptability of the application based on ISO 9126 software quality and the level of satisfaction for its major features. For the statistical treatment of the data collected, Likert Scale, weighted mean and *t*-test were utilized by the researcher.

*Results* – The results revealed that instructors find the existing way of checking attendance as time consuming and a tedious task. Furthermore, the respondents assessed the developed application as moderately acceptable in terms of functionality, reliability and usability while portability was rated as highly acceptable. With regards to the features, the respondents were very satisfied. The t-test also revealed that there is a significant difference between the level respondents' level of acceptability on the existing and the proposed system.

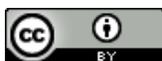

*Conclusion* – The researcher concluded that the developed application was useful and it can support the needs of the instructors to make attendance checking and monitoring easier, faster, and reliable.

*Recommendations* – Due to its acceptable evaluation result, instructors should consider the use of this tool as an alternative to the existing process of checking and monitoring class attendance**.**  The developed system can be enhanced in terms of user design to make it more user-friendly.

*Research Implications* – Classroom presence is important because this helps students to succeed in their academics when they attend class regularly. With the integration of different technologies such as Android, face recognition and SMS, the traditional way of checking class attendance can be made easier, faster, reliable and secured, thus improving classroom management.

## INTRODUCTION

Class attendance is one of the most important factors that help students to complete their understanding and to improve their knowledge and skills.  Attending class regularly can help students to retain information longer, especially when they are participating in class activities, and sharing their opinions and working in groups. These activities enhance the students' ability to keep knowledge in their mind. Moreover, students attending class regularly enhances their necessary skills. By attentive listening and positive contribution in class, skills in speaking, listening, and presentation will be improved significantly. Students can control unexceptional situations not only in their studies but also in the real life. Hence, class attendance is positively and significantly related to student performance (Lukkarinen, Koivukangas, & Seppala, 2016).

In the traditional classroom setting, the existing system or manual way of checking attendance is a roll call done by the teacher where students normally raise their hands or answer "Present!" when called for class attendance. For a lecture class, this process consumes time and effort. Another way of checking attendance is students are asked to write their names and signatures in attendance sheets. The problem with this approach is the authenticity of the attendance records since a student can easily write the name of his/her classmates since this attendance sheet is submitted after class. Moreover, the traditional mode of attendance management leads to a lot of paperwork and it is hard to maintain this attendance sheets for a long period. Since this involved manual recording of attendance, it is prone to inconsistency in terms of data entry of attendance records and it is hard to generate attendance reports. There are also instances that attendance records are kept in computers, but the teachers have to check these files to track student attendance records such as number of absences and tardiness.

Nevertheless, the use of technology in checking class attendance has become easy and convenient. Different innovative technologies and processes have been utilized to eliminate the burden of checking class attendance manually. Some of these technologies include RFID, QR codes, bar codes and biometric systems. Among these technologies, the use of biometric systems offers the greatest amount of authentication.

Techopedia (2018) defines a biometric system as a technological system that uses information about a person to identify that person. Biometric systems rely on specific data about unique biological traits in order to work effectively. A biometric system will involve running data through algorithms for a particular result, usually related to a positive identification of a user or other individual. There are different types of biometric systems from fingerprint, iris, vein, palm, voice, signature, and face recognition.

Face Recognition is one way of authenticating a person using a camera or any image capturing device. In definition, it is a computer application that can identify and verify a person from a video source. It can verify by means of comparing selected features to image files from a face database. Face recognition has been used in many projects and systems around the world.

In addition to biometric systems, the use of SMS technology, also known as "text messaging" has been used in attendance monitoring. With SMS technology, message about students attendance are being sent to parents or guardians for them to be aware that students are attending their respective subjects or classes. This method also ensures the safety of students.

With this, the researcher aimed to develop a class attendance monitoring system using face recognition and SMS technology for the College of Engineering and Computing Sciences (CECS) that will addressed the different challenges encountered on the traditional way of checking attendance. The developed system also intended to improve the way of checking and monitoring of attendance making it easier and faster. With the use of android phones, teachers can easily check class attendance and monitor attendance records while students can easily view their attendance details. Attendance grades are also automatically computed.

**LITERATURE REVIEW**

This section describes previous works done for attendance monitoring. Classroom presence is important because this helps students to succeed in their academics when they attend class regularly. Different researches have shown that a student's regular attendance maybe one of the greatest factor influencing their academic success. With this, different attendance monitoring systems, from manual to automated ones have been utilized to ensure that student attendance are checked and monitored without failed.

In the study of Mendonca et al. (2015), they designed an online attendance system which reduces the duration of the entire attendance checking process. It provide an easier and faster way of attending checking by replacing the traditional process wherein the teachers has to roll call every student name in class and mark the attendance when the student responded. In their proposed system, teachers will no longer need to carry a sheet of paper to mark the attendance and they can also generate attendance records by retrieving the needed data from database, thus, making the entire process paperless.

Islam et al. (2017) proposed an attendance system which uses smart phones. Using the system installed on the smartphones, teachers can easily check attendance and attendance records are saved on the phone SQLite database as well as in the MySQL database simultaneously. The MySQL database is located on the Web server. In addition to attendance checking, the system is capable of calculating attendance percentage, print attendance records, and send email and SMS to the guardian to keep them updated about their child's attendance at the school. If a student calculated percentage of attendance is less than the required percentage, an email is sent to the student's guardian including percentage details as well as a warning. In the parent is not checking their email, an SMS is sent. Since the system is online, it can accessed from any place and any moment which helps the teacher to keep track of their student attendance,

Another study was designed and implemented to utilize mobile devices in attendance management system. An android based mobile attendance management system was developed using VB.NET and SQL Server. Through this project, student attendance can be maintained, student attendance marks are calculated and it also provides a report generation module. The system consists of five modules namely: admin module, registration module, student module and SMS and Android module. Android module allows students to send messages to the system to inform teachers the reason for their leave/absence. Parents can also receive SMS about student attendance (Somasundaram, Kannan, & Sriram, 2016).

An attendance management system was developed to provide solutions to lecture attendance problems using Android devices. In order to use the system, both teachers and students have to install apk files on their devices and were given unique ID and password. Students need to enter their details in the application along with their parent details. During attendance checking, the teacher needs to activate the application on the server and when the application is active, the student can mark the attendance with just one click. Teachers can generate attendance reports weekly as well as monthly. At the end of each month, SMS are sent to parents/guardians of the students, thus informing them about their attendance (Kumbhar et al., 2014).

Likewise, Jabbar Hameed (2017) designed and implemented a smart student attendance system based on Android operating system. The system provides faster, cheaper, and reachable system for online student attendance and generate attendance report automatically. The attendance system has three parts: the admin account that can login to the system and edit the database, the instructor account, which logs to the system to mark student attendance and the reporter who logs to the system to check attendance records and reports all tasks.

Nithya (2015) developed an attendance system that used face recognition using personal component analysis (PCA) algorithm to identify a person. Using the face recognition module, individual student's face is recognized, and attendance is recorded to the database automatically. In addition, student attendance details are sent to the school staff and the parent using e-mail.

In addition, Arun et al. (2014), also utilize face detection and recognition on the system they proposed. The system consists of a camera that captures the images of the classroom and sends it to the image module for enhancement. After the image enhancement, it will proceed with the face detection, where face will be detected from the picture and then face recognition will be used to recognize whose faces get detected and then it will be marked on the database. Pictures of students are stored on the database by the time the students got enrolled.

Varadharajan et al. (2016) also described face recognition technology on their study. In this method, they fixed a camera inside the classroom to capture images. Once faces are detected and recognized with the database, the attendance is marked as present. If the attendance is marked as absent the message about the student's absence is send to their parents.

Previous works mentioned served as a guide in the development of the application. Some features of the previous systems were adapted such the integration of mobile phones, face recognition, and SMS technology. Since majority of the population today has Android phones, the researcher came up with the idea of developing an attendance system that is portable and can be accessed anytime anywhere. With the use of Android phones, the teacher can easily check attendance without bringing paper and laptops. For the students, they can view attendance details without difficulty using their Android devices. Moreover, text messaging or SMS technology was used to inform parents/ guardians about their child's attendance and to guarantee the safety of students. In order to have an authentic attendance records, face recognition technology was utilize. Using the camera of the Android phone, individual student's face is recognized and mark as present on the database. Attendance reports can also be generated when needed.

## METHODOLOGY

### Software Development

The researcher utilized the incremental model as the methodology in developing the application. The incremental model, as shown on Figure 1, is a method of software development where the model is designed, implemented and tested incrementally until the product satisfies all of its requirements. According to Ghahrai (2018), this model combines the elements of the waterfall model with the iterative philosophy of prototyping methodology.

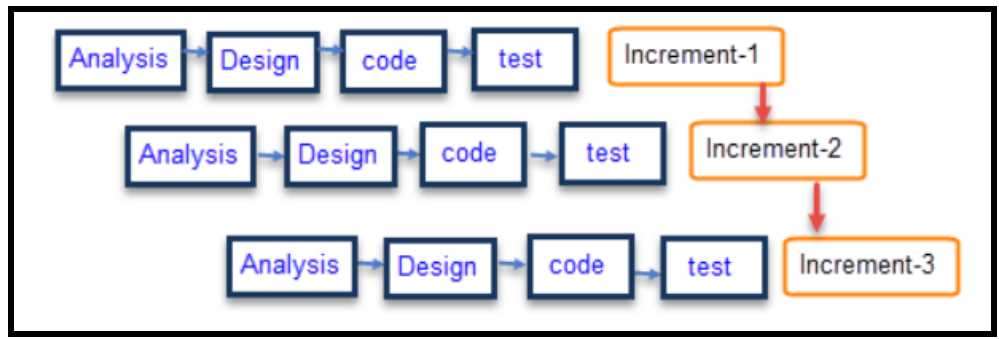

*Figure 1.* The Incremental Model

On the requirement analysis phase, the researcher defined the requirement and specification of the application. In incremental model, the requirements of the application must be clearly understood. First, the researcher developed a questionnaire that discusses the different challenges/problems encountered on the existing system of checking attendance. These challenges will be addressed by the features to be integrated on the application. When the requirements and specification are defined, the design phase followed. This phase involves the design of the user interface, forms and the database. After the design phase, the researcher proceeds with the actual coding of the application to integrate the expected features of the system. Once the coding phase is completed, the application was tested to point out errors on the development process and to ensure the quality of the developed application.

The initial version of the application was examined by selected CECS faculty members and their comments and suggestions were considered on the next version of the application. The next iteration passed through the same phases: requirement analysis, design, code and test. The development of the application was completed when all the designed features have been integrated.

## Conceptual Framework

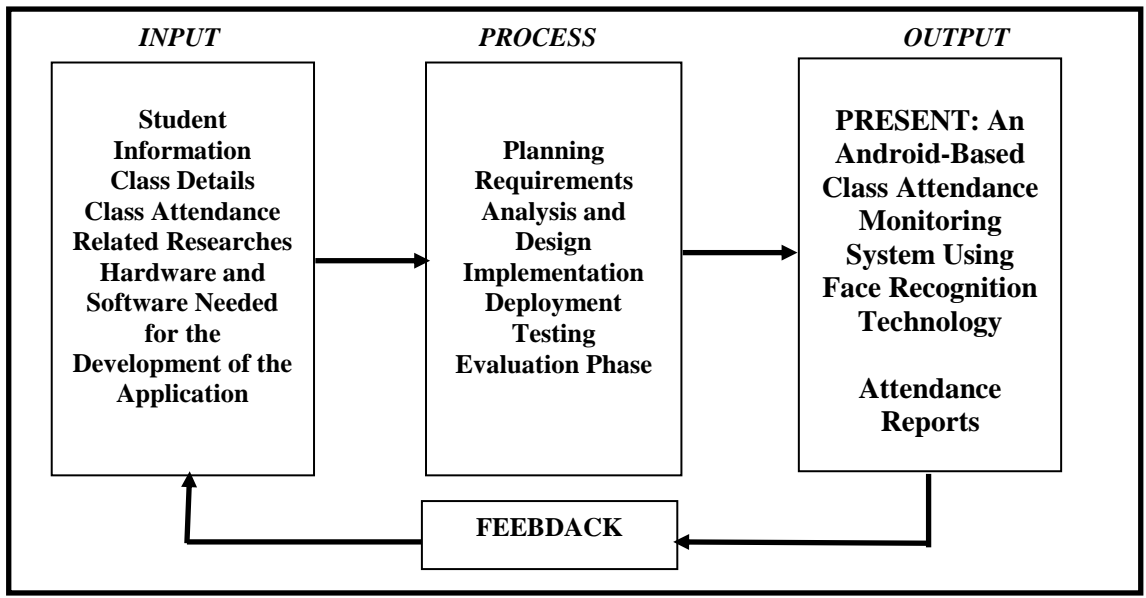

*Figure 2.* Input-Process-Output Chart

Figure 2 describes the conceptual framework of the research study. Input refers to all the external data needed for the completion of the study. These include student information, class details, class attendance and the hardware and software needed to develop the application. Related researches have also been reviewed to enhance and deepen the knowledge about the research study.

Process is the most significant element of the study. This involves the different phases that the researcher has undertaken to develop the system. After the development phase, it will undergo evaluation phase by using a self-constructed evaluation questionnaire to determine the level of acceptability and satisfaction of the respondents on the developed system. Different statistical tools were also used to compute and interpret the gathered data.

The output frame displays the result after processing the input. This refers to the developed application along with the generated outputs such as attendance reports and attendance grades.

## Hardware and Software Needed for the Development

Listed on Table 1 and Table 2 are the hardware and software used for the development of the application.

Table 1. Software Requirements

| Software | Specification |
|---|---|
| Operating System | 32 bit or higher |
| Web Editor | Xampp |
| IDE | Android Studio |
| SDK | JDK |
| Database | SQLite |
| Android | Gingerbread or Higher |

Table 2. Hardware Requirements

| Android Phone | | |
|---|---|---|
| | RAM | 1GB or higher |
| | Camera | Front 13-megapixels |

## System Testing and Evaluation Phase

After the development of the application, it underwent an evaluation phase. The researchers constructed a self –made evaluation questionnaire, as shown on Table 3, which evaluates the developed application in terms of software quality such as functionality, reliability, usability and portability based on ISO 9126. The existing system of checking and monitoring attendance was also assessed in order to test the significant difference of the existing system and proposed application. In addition, the vital features of the application were also evaluated. A scale of 1 to 5 was used, with 5 as the highest and 1 as the lowest. Seventeen (17) faculty members from the College of Engineering and Computing Sciences (CECS) at Batangas State University ARASOF Nasugbu Campus were asked to evaluate developed application.

After the evaluation phase, the answer from the evaluation questionnaire was tallied and tabulated. For the statistical treatment of data, frequency count, weighted mean and t-test was used. Using Likert scale, the average weighted mean was computed and interpreted with a corresponding verbal interpretation (Table 4).

One of the objectives of the study is to test if there is a significant difference between the level of acceptability of the existing and proposed system in terms of stated software quality. A null and alternative hypothesis was formulated.

$H_o$: *There is no significant difference on the respondent's level of acceptability on the existing and proposed system.*

*Ha: There is a significant difference on the respondent's level of acceptability on the existing and proposed system.*

Table 3. Components of the Evaluation Questionnaire

| Software Quality | Indicators |
|---|---|
| Functionality | Performs its intended specification |
|  | Carry out its function w/ less time and effort |
|  | Supports the needs of every user |
| Reliability | Reliable to use at all time |
|  | Produces reliable and accurate attendance information |
|  | Matches its stated specification without failure |
| Usability | Involves processes that is easy to learn and understand |
|  | Easy to use and operate |
|  | Usable to every users |
| Portability | Available to be used anytime and anywhere. |
|  | Produces output that can be accessed at all time |
|  | Generates output/result in a portable format |

| Features | Indicators |
|---|---|
| Checking and Monitoring of Attendance | Can check and record student attendance using face recognition |
|  | Can monitor student attendance easier and faster.(tardiness, number of absences) |
|  | Can display correct and authentic attendance records |
| Attendance Grade Computation | Can compute attendance grade automatically |
|  | Can help user to be aware of their attendance records and grade |

Table 4. Likert Scale for the Level of Acceptability and Satisfaction

| Average Weighted Mean Range | Descriptive Equivalent (Level of Acceptability) | Descriptive Equivalent (Level of Satisfaction) |
|---|---|---|
| 4.21 – 5.00 | Highly Acceptable | Extremely Satisfied |
| 3.41 – 4.20 | Moderately Acceptable | Very Satisfied |
| 2.61 – 3.40 | Acceptable | Satisfied |
| 1.80 – 2.60 | Fairly Acceptable | Slightly Satisfied |
| 1.00 – 1.79 | Poorly Acceptable | Not Satisfied |

# RESULTS AND DISCUSSIONS

On the requirements analysis phase, a survey questionnaire was constructed to determine the different challenges/problems encountered by the CECS faculty members in checking and monitoring of class attendance. The data gathered will be used to determine the important features and functionalities needed on the developed application to address the challenges/problems identified.

Table 5. Ranking of the Common Problems/Challenges Encountered in Manual Checking and Monitoring of Attendance

| Method | Frequency | Rank |
|---|---|---|
| Checking of attendance is time consuming. | 14 | 1 |
| Monitoring attendance is tedious task (tracking of number of absences and tardiness of students) | 11 | 2 |
| Attendance records being misplaced or lost. | 3 | 5 |
| Computation of attendance grade involves a lot of effort. | 8 | 4 |
| Attendance record is not that accurate. | 7 | 3 |

Table 5 displays the different problems encountered on the existing system of checking and monitoring of attendance. It shows that majority of the respondents agreed that manual way of checking attendance is time consuming, followed by monitoring attendance is a tedious task. On the last rank, attendance records are misplaced or lost.

Table 6. Level of Acceptability of the Existing System

| Software Quality | Weighted Mean | Verbal Interpretation |
|---|---|---|
| Functionality | 3.53 | Moderately Acceptable |
| Reliability | 3.49 | Moderately Acceptable |
| Usability | 368 | Moderately Acceptable |
| Portability | 3.45 | Moderately Acceptable |
| **Average Weighted Mean** | **3.54** | |

Table 6 shows that the respondents rated the existing system as *Moderately Acceptable* based on functionality, reliability, usability and portability. This result means that the existing system needs improvement to address the different problems/challenges encountered. The average weighted mean is 3.54.

Table 7. Level of Acceptability of Developed Application

| Software Quality | Weighted Mean | Verbal Interpretation |
|---|---|---|
| Functionality | 4.12 | Moderately Acceptable |
| Reliability | 4.10 | Moderately Acceptable |
| Usability | 4.10 | Moderately Acceptable |
| Portability | 4.27 | Highly Acceptable |
| **Average Weighted Mean** | **4.17** | |

Based on Table 7, functionality, reliability, and usability was rated as Moderately Acceptable. On the other hand, portability is Highly Acceptable. With this high evaluation rating, the developed system was proved to be useful and meets the needs of the respondents.

Table 8. Level of Satisfaction of the Features of the Developed Application

| Features | Weighted Mean | Verbal Interpretation |
|---|---|---|
| Checking and Monitoring Class Attendance | 4.14 | Very Satisfied |
| Automatic Attendance Grade Computation | 4.16 | Very Satisfied |

As indicated on Table 8, the respondents rated the features of the developed application as *Very Satisfied*. The evaluation rating can be attributed that the developed system can performs its stated features easier and faster and helps to improve classroom management.

Table 9. Significant Difference between the Manual System and the Developed Application

| Groups | Weighted Mean | $t_c$ (computed value of *t*-test) | $t_{0.05(22)}$ (table value of *t*-test) |
|---|---|---|---|
| Existing System's Mean on the Level of Acceptability | 3.54 | 7.58 | 1.72 |
| Proposed System's Mean on the Level of Acceptability | 4.17 | | |

Table 9 shows the significant difference of the manual system and the developed system using T-test.

Decision Rule: The negative sign of the computed value of *t*-test implies that the existing system's mean is less than the mean of the proposed system's mean. The tabular value of *t* for df(22) at 0.05 level is *1.72*. Since the computed value of *t*-test does exceed the critical value, the researchers reject the null hypothesis. *There is a significant difference between the level respondents' level of acceptability on the existing and the proposed system.*

**Sample Screenshots**

The following figures below displays the main screen shots for the application.

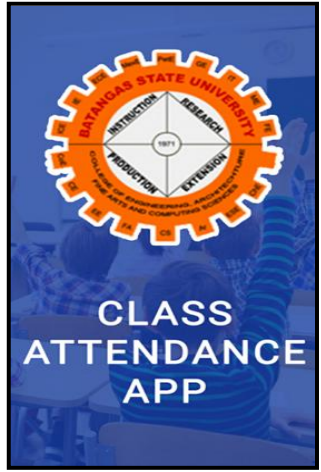 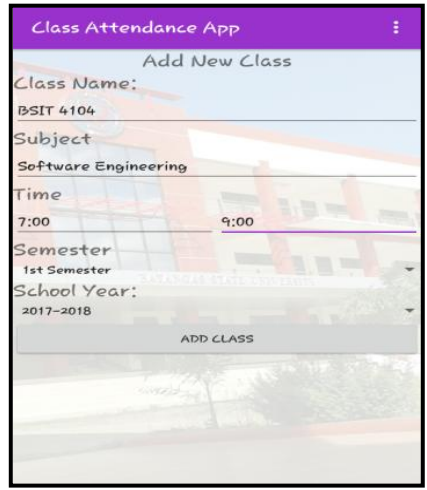

*Figure 3.* The Main Interface of the Application    *Figure 4.* Add New Class Interface

Figure 3 displays the main interface of the application when it is launch in the Android device. On Figure 4, the Add New Class interface is shown. Using this interface, the teacher can create classes, add class details and then register students.

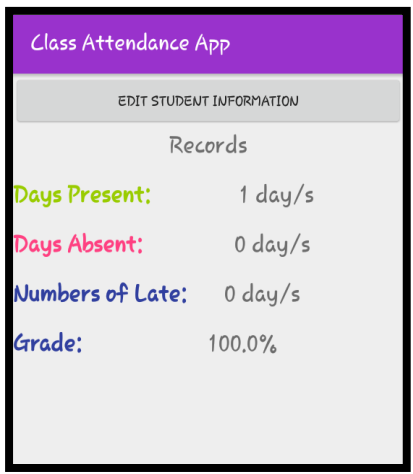

*Figure 5.* Attendance Grade Computation   *Figure 6.* Attendance Record in Excel Format

Figure 5 shows that the student can view his/her the attendance records such as the absences and tardiness. The application can also automatically compute the attendance grades. Figure 6 displays that the instructor can export the attendance records in MS Excel format for report generation.

## CONCLUSIONS AND RECOMMENDATIONS

From the findings of the study, the following conclusions were drawn:

1. The existing system cannot completely support the needs of the instructors in terms of checking and monitoring attendance. The user wants to have a new way of checking and monitoring of attendance.
2. The existing system needs improvement to make it more functional, reliable, usable, and portable on the part of the users.
3. Based on the evaluation of the respondents, the android application was confirmed more useful, complete and accepted than the existing system due to its respectable rating. In addition, the evaluation result also shows the openness of the respondents for adopting a new system that will improve classroom management.
4. The evaluation of the respondents proved that the developed system is a good alternative to the existing system to make attendance checking easier, faster, and reliable.

Due to its acceptable evaluation result and the respondents are satisfied with features of the developed application, instructors should consider the use of this tool as an alternative to the existing process of checking and monitoring class attendance. The developed system can be enhanced in terms of user design to make it more user friendly.

## IMPLICATIONS

Classroom presence is important because this helps students to succeed in their academics when they attend class regularly. With the integration of different technologies such as Android, face recognition and SMS, the traditional way of checking class attendance can be made easier, faster, reliable and secured, thus improving classroom management.

## ACKNOWLEDGEMENT

This is to acknowledge the contributions of my institution, mentors, family, colleagues, friends and students who has been a constant source of love, concern, support and strength all these years. And to my husband and best friend, Renz, for his love, encouragement, and continuous moral support which motivated me to remain focused towards achieving various milestones of my journey.